\documentclass[sigconf,table]{acmart}
\AtBeginDocument{%
  }

\usepackage{paralist}

\acmYear{2025}
\setcopyright{cc}
\setcctype{by}
\acmConference[ICTIR '25]{Proceedings of the 2025 International ACM SIGIR Conference on Innovative Concepts and Theories in Information Retrieval (ICTIR)}{July 18, 2025}{Padua, Italy}
\acmBooktitle{Proceedings of the 2025 International ACM SIGIR Conference on Innovative Concepts and Theories in Information Retrieval (ICTIR) (ICTIR '25), July 18, 2025, Padua, Italy}\acmDOI{10.1145/3731120.3744617}
\acmISBN{979-8-4007-1861-8/2025/07}
\begin{document}


\title{Answering Multimodal Exclusion Queries with Lightweight Sparse Disentangled Representations}

\author{Prachi J}\authornote{Work done as an intern at the Media and Data Science Research Lab, Adobe Systems}
 \affiliation{
  \institution{Indian Institute of Technology, Delhi}
   \city{New Delhi}
   \country{India}}
 \email{prachi@cse.iitd.ac.in}

\author{Sumit Bhatia}
 \affiliation{
 \institution{Media and Data Science Research Lab, Adobe Systems}
   \country{India}}
 \email{sumit.bhatia@adobe.com}

\author{Srikanta Bedathur}
 \affiliation{
  \institution{Indian Institute of Technology, Delhi}
  \city{New Delhi}
  \country{India}}
 \email{srikanta@cse.iitd.ac.in}
\renewcommand{\shortauthors}{Prachi et al.}

\ccsdesc[500]{Information systems~Image search}
\ccsdesc[500]{Computing methodologies~Neural networks}
\keywords{Disentanglement, Multimodal, exclusion based query}


\begin{abstract}
Multimodal representations that enable cross-modal retrieval are widely used. However, these often lack interpretability making it difficult to explain the retrieved results. Solutions such as learning sparse disentangled representations are typically guided by the text tokens in the data, making the dimensionality of the resulting embeddings very high. We propose an approach that generates smaller dimensionality fixed-size embeddings that are not only disentangled but also offer better control for retrieval tasks. We demonstrate their utility using challenging \emph{exclusion} queries over MSCOCO and Conceptual Captions benchmarks. Our experiments show that our approach is superior to traditional dense models such as CLIP, BLIP and VISTA (gains up to 11\% in AP@10), as well as sparse disentangled models like VDR (gains up to 21\% in AP@10). We also present qualitative results to further underline the interpretability of disentangled representations.

\end{abstract}

\maketitle
  \section{Introduction}\label{introduction}
\begin{figure}[t]
\centering
\includegraphics[width=0.9\columnwidth]{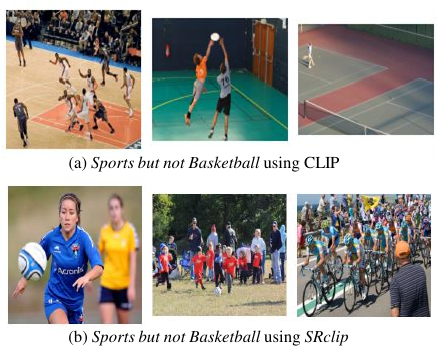} 
\caption{Top results for \textit{exclusion} query using CLIP and proposed \textit{SR}\textsubscript{\textit{clip}} embeddings.}
\label{Fig:fig1}
\end{figure}

Multimodal representations help to integrate and process information from different types of data, or modalities, such as text and image and have shown great application in many downstream tasks, including multimodal retrieval. However, these models suffer from poor intrepretability, making it challenging to fully understand and explain how they combine different types of information. 
Disentanglement~\cite{bengio2014representationlearningreviewnew, wang2024disentangled, chen2019isolatingsourcesdisentanglementvariational, higgins2018definitiondisentangledrepresentations} addresses some of these challenges by separating the various underlying \textit{factors of variation} within the data, and thus, enhancing the explainability, interpretability, controllability, and generalizability of the representations~\cite{bengio2014representationlearningreviewnew,Greff2020OnTB,wang2024disentangled}.
The key challenge in disentanglement is identifying these factors of variation.  Early research efforts, such as $\beta$-VAE~\cite{higgins2017betavae}, FactorVAE~\cite{kim2019disentanglingfactorising},  and Relevance FactorVAE~\cite{kim2019relevancefactorvaelearning}, primarily focused on synthetic datasets like Shapes3D~\cite{3dshapes18}, where predefined and well-structured factors enabled direct evaluation. While some studies have extended disentanglement techniques to multimodal settings~\cite{article,lee2020privateshareddisentangledmultimodalvae,Kim2021LearningDF}, they are often restricted to synthetic or relatively simple real-world datasets with a fixed and limited number of factors of variation.

\begin{figure*}[t]
\centering
\includegraphics[width=\textwidth]{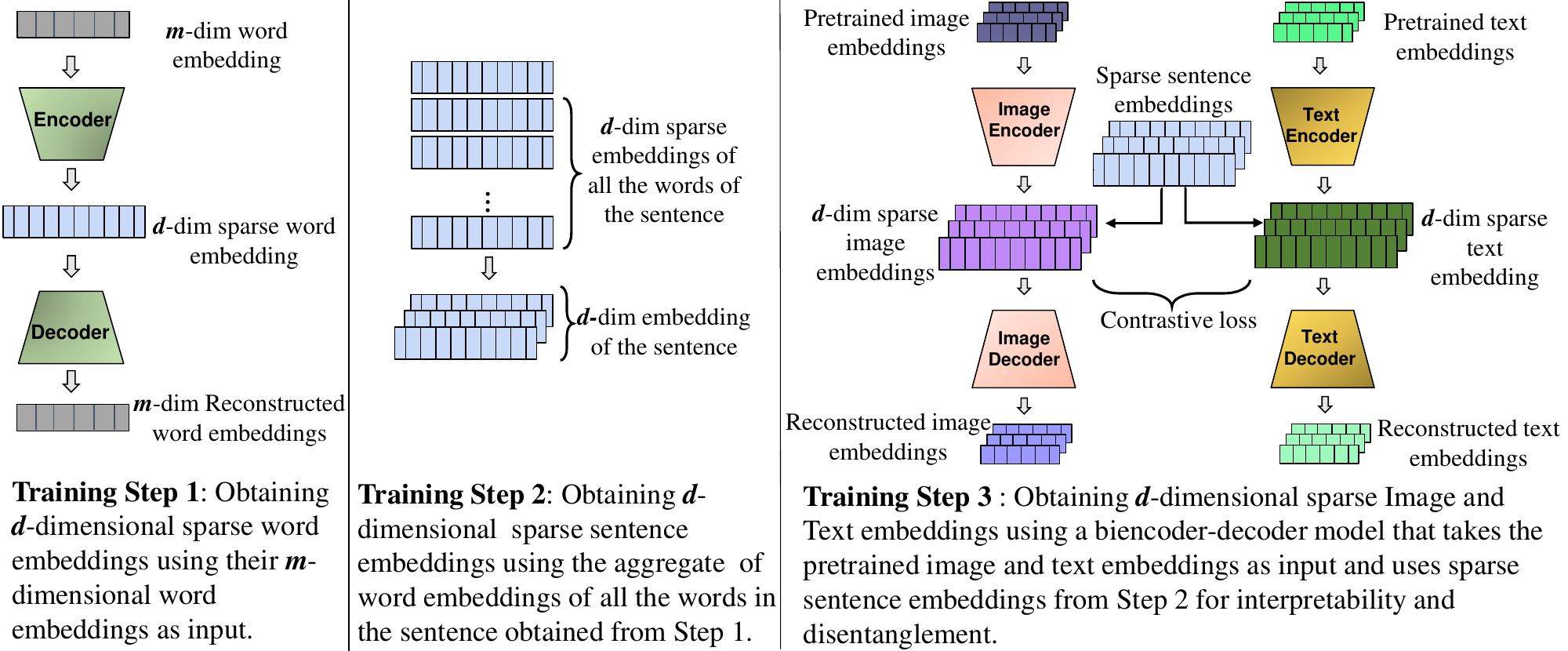} 
\caption{{Illustration of the 3-step training process for generating sparse, disentangled representations. Training step 1 produces interpretable word embeddings for all the words in the vocabulary, which are then used in the second step to create sentence embeddings. In training step 3, a biencoder decoder model is used to create sparse disentangled embeddings of images and texts by using the sentence embeddings created in Step 2 as a guiding bias to activate particular dimensions.}}
\label{Fig:fig2}
\vspace{-10px}
\end{figure*}

However, in complex real-world multimodal datasets, where the number of factors is not predetermined, disentangling representations becomes significantly more challenging ~\cite{Bhalla2024InterpretingCW, zhou2024retrievalbaseddisentangledrepresentationlearning}. One promising way is to leverage the vocabulary of the associated text to capture different factors -- each token corresponds to one unique factor. Vocabulary Disentangled Retrieval (VDR)~\cite{zhou2024retrievalbaseddisentangledrepresentationlearning} uses this approach and maps each word or token in the vocabulary to a single dimension in the representation. While this approach can capture a large variety of factors, it leads to prohibitively large representations (embedding dimensions = vocabulary size). To address this issue, we propose a model that captures key factors from textual captions using significantly more compact representations by employing a simple intuition --- instead of assigning each word or token its own dimension, we use subsets of dimensions to represent similar words and concepts. Given that there are $2^N-1$ proper subsets for a set of size $N$, even a moderate embedding dimensionality of $1000$ can capture practically all factors of interest. Thus, our proposal produces disentangled representations by separating concepts based on dimension \textit{subsets} 
and enables more efficient handling of real-world data with compact, sparse embeddings.

Once disentangled embeddings are obtained, effectively separating factors and components within the dataset, they can be manipulated by selectively excluding specific components through adjustments in the corresponding dimensions. This capability is particularly valuable in retrieval scenarios, where users may need not only to search for specific content but also to explicitly exclude certain elements. Handling exclusion is essential for precise retrieval but remains a significant challenge for current models for various retrieval tasks, including document-retrieval~\cite{Ravichander2022CONDAQAAC,zhang2024excluirexclusionaryneuralinformation}, image-retrieval~\cite{10.1145/3652583.3657619}, and multi-modal retrieval~\cite{bui2024neintellingdontwant, singh2024learnnosay}. Existing multimodal representation models, such as CLIP~\cite{radford2021learningtransferablevisualmodels} and BLIP~\cite{li2022blipbootstrappinglanguageimagepretraining}, as well as state-of-the-art multi-modal retrieval approaches like VISTA~\cite{zhou-etal-2024-vista}, struggle to accurately interpret negation queries. By leveraging disentangled representations, models can enhance their ability to handle exclusion, improving their capacity to interpret and act upon negation-based queries.

To illustrate exclusion-based retrieval, we present an example (Figure~\ref{Fig:fig1}) comparing image retrieval results using CLIP and our proposed \textbf{\textit{Sparse Representation}} of Clip ($SR_{clip}$) for the exclusion query \emph{sports but not basketball}. Figure~\ref{Fig:fig1}(a) shows the retrieval results for the query using CLIP embeddings, where basketball-related images are still present, highlighting CLIP’s inability to handle negation effectively. In contrast, Figure~\ref{Fig:fig1}(b) presents the retrieval results using $SR_{clip}$, where basketball images are successfully excluded, demonstrating the effectiveness of our approach in exclusion-based retrieval.

\noindent
\textbf{Our Contributions:} The key contributions of this work are as follows:
\begin{enumerate}[1)]
\item We propose a novel method for disentangling factors of variation in multimodal data while significantly reducing embedding dimensionality compared to conventional approaches. 
\item We introduce a retrieval framework specifically designed to better handle exclusion-based queries.
\item We release a new dataset for evaluating exclusion in multi-modal retrieval tasks and benchmark the performance of our proposed method against various state-of-the-art baselines. Our code and dataset are available \href{https://anonymous.4open.science/r/retrieval25-C485/}{\url{https://anonymous.4open.science/r/retrieval25-C485/}}.
\end{enumerate}

  \section{Related Work}
\noindent
\textbf{Disentangled Representation} aims to separate different factors of variation in data into different dimensions or vectors. This allows models to reason about individual semantic attributes and enables better control, generalization, and explainability~\cite{Greff2020OnTB,wang2024disentangledrepresentationlearning}. Early approaches focused primarily on image data, especially synthetic datasets with fixed factors of variation. Variants of Variational Autoencoders (VAEs), including $\beta$-VAE~\cite{higgins2017betavae}, FactorVAE~\cite{kim2019disentanglingfactorising}, and Relevance Factor VAE~\cite{kim2019relevancefactorvaelearning}, attempted to enforce axis-aligned latent dimensions such that each encoded a single interpretable factor like shape, position, or color. However, these methods were mostly limited to simplified datasets, and their ability to scale to real-world data was limited. Extending disentanglement to multimodal settings introduces additional challenges—not only must latent factors be separated within each modality, but semantically aligned across modalities. Recent works like~\cite{article,lee2020privateshareddisentangledmultimodalvae,Kim2021LearningDF} address this by disentangling modality-specific and modality-invariant components, enabling joint reasoning across image-text or audio-visual pairs. However, unlike unimodal settings where factors are explicitly aligned to dimensions, multimodal disentanglement often focuses on latent space partitioning, making interpretability less direct. Disentangled representations have also been leveraged for task-specific applications across diverse domains. In recommendation systems, disentanglement helps isolate user preferences or intents, such as in MacridVAE~\cite{ma2019learningdisentangledrepresentationsrecommendation} and IEDR~\cite{Su_2025}, which decouple intrinsic and extrinsic factors of decision-making. In scene understanding and image translation tasks, disentanglement enables reasoning under occlusions, for example, by separating body parts or background clutter in person re-identification~\cite{10.1109/TMM.2022.3141267}, or disentangling rain streaks from scene content~\cite{pizzati2020modelbasedocclusiondisentanglementimagetoimage}. Similarly, in scene text processing, style-content disentanglement improves text recognition and editing~\cite{zhang2024chooseneeddisentangledrepresentation}. In retrieval and alignment tasks, sparse disentangled representations have shown promise in making representations both interpretable and efficient. SpLice~\cite{Bhalla2024InterpretingCW} and VDR~\cite{zhou2024retrievalbaseddisentangledrepresentationlearning} extend disentanglement to real-world multimodal data, using sparse embeddings to represent high-level semantic factors. VDR, in particular, treats natural language as a supervisory signal to construct a shared vocabulary space where aligned image-text pairs activate semantically meaningful dimensions—such as specific objects or attributes—enabling compositional and interpretable retrieval. Complementary to these efforts, other works have explored sparsity as a standalone tool for efficient and interpretable retrieval. LexiLip~\cite{Luo_2023_ICCV}   introduces a lexicon bottleneck to constrain image-text alignments via predefined concept vocabularies, while VisualSparta~\cite{lu-etal-2021-visualsparta} uses fragment-level matching with sparse attention to accelerate retrieval. More recently, ~\cite{nguyen2024multimodallearnedsparseretrieval} propose a lightweight method that converts dense multimodal vectors into sparse lexical representations using a small projection head and probabilistic expansion control, focusing on efficiency and practical integration. However, these models prioritize speed and scalability rather than explicitly encoding disentangled factors. Sparsity has also played a central role in text representation learning, where models like SPINE~\cite{Subramanian2017SPINESI} pioneered sparse interpretable embeddings for NLP tasks. Subsequent works such as SparTerm~\cite{bai2020spartermlearningtermbasedsparse}, SPLADE~\cite{formal2021spladesparselexicalexpansion}, and SPLADEv2~\cite{formal2021spladev2sparselexical} built on this foundation with sparse lexical expansion techniques, improving both retrieval effectiveness and model transparency. These approaches reinforce the synergy between sparsity and disentanglement—sparse activations often lead to more interpretable latent dimensions by isolating specific semantic concepts.
\\ \noindent
\textbf{Exclusion-Based Retrieval}, is an emerging paradigm that focuses on queries containing both inclusion and exclusion criteria—retrieving results that include certain concepts while explicitly omitting others. In the text domain, several recent studies have highlighted the complexity of handling negation and exclusion. ExcluIR~\cite{zhang2024excluirexclusionaryneuralinformation} introduced the first large-scale benchmark and training dataset specifically designed for exclusion-aware retrieval, enabling systematic evaluation of models on exclusion-based queries. NevIR~\cite{weller-etal-2024-nevir} proposed a contrastive benchmark that tests a model’s sensitivity to negation by contrasting documents differing only in negated terms, revealing that many current systems fail to account for such cues effectively. Similarly, \cite{garcia-ferrero-etal-2023-dataset} presented a large benchmark evaluating large language models across diverse forms of negation, demonstrating that even state-of-the-art models often struggle to generalize beyond surface-level patterns. \cite{malaviya-etal-2023-quest} introduced the Quest dataset, which maps queries with logical constraints (e.g., negation, conjunction) to Wikipedia entities, requiring models to reason over document evidence and perform set operations—an objective closely aligned with the exclusion-based evaluation dataset we propose for the multimodal domain.

In the multimodal domain, exclusion-based retrieval is relatively underexplored. One of the earliest efforts is presented by~\cite{10.1145/3652583.3657619}, who propose a method for exclusion-based image search. Recent works such as~\cite{bui2024neintellingdontwant, singh2024learnnosay} have extended exclusion handling to multimodal generation and retrieval tasks. In particular, \cite{singh2024learnnosay} introduced the CC-Neg dataset, which consists of exclusion queries derived from the Conceptual Captions dataset. However, it lacks ground-truth image annotations for those queries, limiting its utility for evaluating retrieval performance. As retrieval systems evolve to support fine-grained control through constraints like exclusion, these studies underscore the growing need for exclusion-aware modeling. Nevertheless, exclusion handling—especially in the multimodal setting—remains an underexplored yet promising direction for future research.

  \begin{table*}[t]
\centering
\caption{\small Results for various methods on MSCOCO and Conceptual Captions Datasets for Exclusion Based Retrieval. We report numbers for SLQ, Avg. Emb. and SR methods with both CLIP and BLIP as base representation models. Statistically significant improvements over VDR, SpLice, Content-Based Exclusion, VISTA, CLIP SLQ, CLIP Avg Emb, BLIP SLQ and BLIP Avg Emb are indicated by superscripts 0, 1, 2, 3, 4, 5, 6 and 7, respectively(measured by paired t-Test with 99\% confidence)}
\label{combined_table1}
\resizebox{\textwidth}{!}{%
\begin{tabular}{@{}lllllllll@{}}
\toprule
             & \multicolumn{4}{c}{\textbf{MSCOCO}}   & \multicolumn{4}{c}{\textbf{Conceptual Captions}}    \\ 
    \cmidrule(r){1-1}
    \cmidrule(rl){2-5}
    \cmidrule(rl){6-9}
    \textbf{Method} & \textbf{MRR@1} & \textbf{MRR@10} & \textbf{NDCG@10} & \textbf{AP@10} & \textbf{MRR@1} & \textbf{MRR@10} & \textbf{NDCG@10} & \textbf{AP@10} \\
    \cmidrule(r){1-1}
    \cmidrule(rl){2-5}
    \cmidrule(rl){6-9}
                       $\quad$\textbf{VDR}   & 0.7195 & 0.7873  & 0.6648 & 0.6446 & 0.5473 & 0.6687 & 0.5536 & 0.5512 \\
                       $\quad$\textbf{SpLice} & 0.0718 & 0.1184 & 0.0543 & 0.0518 & 0.0616 & 0.1271 & 0.0564 & 0.0553 \\
                       $\quad$\textbf{CBE}    & 0.2960 & 0.4399 & 0.3106  & 0.3114 & 0.2994 & 0.4237 & 0.3027  & 0.3010 \\ 
                       $\quad$\textbf{Vista}   & 0.6212 & 0.7299  & 0.6233 & 0.6191 & 0.4962 & 0.6268 & 0.4758 & 0.4706 \\

                       \multicolumn{3}{@{}l}{\emph{Using CLIP as base model}} &&&&& \\
                       $\quad$ \textbf{SLQ}   & $0.6125^{12}$ & $0.7208^{12}$ & $0.5636^{12}$ & $0.5504^{12}$ & $0.5129^{123}$ &  $0.6325^{123}$ & $0.4759^{123}$ & $0.4658^{12}$ \\
                       $\quad$ \textbf{Avg. Emb.} & $0.7981^{01234}$ & $0.8552^{01234}$ & $0.7293^{01234}$ & $0.7099^{01234}$ & $0.6268^{01234}$ & $0.7375^{01234}$ & $0.6128^{01234}$ & $0.6079^{01234}$ \\
                       \rowcolor{orange!30}
                       $\quad$ \textbf{SR}\textsubscript{clip} & $\underline{0.8669}^{012345}$ & $\underline{0.9175}^{012345}$ & $\underline{0.8064}^{012345}$ & ${0.7865}^{012345}$ & $\underline{0.6749}^{012345}$ & $\textbf{0.7698}^{012345}$ & $\underline{0.6528}^{01234}$ & $\underline{0.6460}^{01234}$ \\ 
                       \multicolumn{3}{@{}l}{\emph{Using BLIP as base model}} &&&&& \\
                        $\quad$ \textbf{SLQ}   &  $0.7117^{1234}$ & $0.8190^{01234}$ & $0.6884^{01234}$  & $0.6768^{01234}$ & $0.5087^{123}$ & $0.6362^{123}$ & $0.4938^{1234}$ & $0.4900^{1234}$ \\
                      $\quad$ \textbf{Avg. Emb.} &  $0.8376^{01236}$ & $0.8815^{01236}$ & $0.7987^{01236}$ & $\underline{0.7868}^{01236}$ & $\textbf{0.7028}^{01236}$  & $\underline{0.7661}^{0125}$ & $\textbf{0.6702}^{0125}$ & $\textbf{0.6620}^{01236}$ \\
                      \rowcolor{cyan!30}
                      $\quad$ \textbf{SR}\textsubscript{blip} & $\textbf{0.9226}^{012367}$ & $\textbf{0.9536}^{012367}$ & $\textbf{0.8553}^{012367}$ & $\textbf{0.8359}^{012367}$ & $0.6290^{01236}$ &  $0.7348^{01236}$ & $0.5820^{01236}$ & $0.5704^{01236}$ \\ \bottomrule 
\end{tabular}%
}
\end{table*}

\section{Methodology}\label{methodology}

We propose a three-step training pipeline (Fig.\ref{Fig:fig2}) to generate sparse, interpretable multimodal embeddings. 
In \textbf{Training Step 1}, we generate sparse and interpretable embeddings for all the words in the vocabulary $\mathcal{V}$ of textual part of the dataset where the aim is to enforce sparsity and interpretability in contextual word embeddings. For this we take $m$ dimensional pretrained word embeddings such as GloVe \cite{pennington-etal-2014-glove} which are projected to $d$ dimensions ($\mathbb{R}^{V \times m} \rightarrow \mathbb{R}^{V \times d}$) using a Sparse Autoencoder\cite{ng2011sparse}. This autoencoder enforces sparsity and creates sparse latent embeddings $\mathbf{z}_{w}$ for the words $\mathbf{w}$, such that semantically similar words have similar dimensions activated. This is achieved using the method explained below, following a similar approach to ~\citet{Subramanian2017SPINESI}. 
\begin{align*}
    \mathbf{z}_w &= f_{\text{enc}}(\mathbf{w}) \\
    \hat{\mathbf{w}} &= f_{\text{dec}}(\mathbf{z}_w)
\end{align*}
where $\mathbf{z}_w \in \mathbb{R}^{d}$ is the sparse latent representation of the word, and $\hat{\mathbf{w}}$ is the reconstructed word embedding. The model is trained using following three loss functions:
\\ \noindent
\textbf{Reconstruction loss(RL)} is the average loss in reconstructing the input representation from learned representation and reconstructed word embeddings.
\begin{equation*}
RL = \frac{1}{\mathcal{V}} \sum_{i \in \mathcal{V}} \left\| \hat{\mathbf{w}}_i - \mathbf{w_i} \right\|_2^2
\end{equation*}
\textbf{Average sparsity Loss(ASL)} penalizes deviations of the observed average activation value from the target activation value for a given hidden unit across a dataset.
\begin{equation*}
ASL = \sum_{h \in H} \max(0, (\rho_{h,\mathcal{V}}-\rho^{*}_{h,\mathcal{V}}))^2
\end{equation*}
where $\rho^{*}_{h,\mathcal{V}}$ is the desired sparsity and $\rho_{h,\mathcal{V}}$ is the actual sparsity calculated by the average of the activations in the hidden or latent layer.\\
\textbf{Partial Sparsity Loss(PSL)} penalizes the values that are neither close to 0 nor 1 and pushes them close to 0 and 1, adding more sparsity to the embeddings.
\begin{equation*}
PSL = \frac{1}{\mathcal{V}}  \sum_{i \in \mathcal{V}} \sum_{h \in \mathcal{H}} (z_{wi}^h * (1-z_{wi}^h))
\end{equation*}
 where $\mathcal{H}$ is set of hidden units in a layer, and the final loss is the sum of RL, ASL and PSL losses. Finally, the $d$-dimensional latent representations $\mathbf{z}_w$ serve as the sparse disentangled word embeddings. The representations obtained using this sparse autoencoder method exhibit inherent interpretability and disentanglement at the dimension level. Similar approach has also been explored in several recent works to create sparse intrepretable embeddings, such as by \citet{huben2024sparse, o'neill2024towards}.\\
\noindent In \textbf{Training step 2}, we compute sentence embeddings for image captions. Given a sentence $c = (w_1, w_2, \dots, w_n)$ with $n$ words, where each word ${w}_{i}$ has a sparse embedding $\mathbf{z}_{w_i}$, the final sentence embedding $\mathbf{z}_c$ is obtained as:
\vspace{-2mm}
\begin{equation*}
z_c = \frac{1}{n} \sum_{i=1}^{n} z_{w_i}
\end{equation*}
These sentence embeddings retain the interpretability of the individual word embeddings while capturing meaningful patterns, with similar words and features having high values in the same set of dimensions.

\noindent In \textbf{Training Step 3}, we use a biencoder-decoder model with paired encoders and decoders for images and text, both sharing the same architecture. The encoders $f_{\text{encoder}}$ take $k$-dimensional pretrained embeddings—$E^{\text{img}}_k$ for images and $E^{\text{text}}_k$ for text—and map them to a $d$-dimensional latent space ($d > k$, $d$ being the same size as in Step 1). The decoders $f_{\text{decoder}}$ then reconstruct the embeddings back to $k$-dimensions, ensuring that the transformed representations retain relevant information.
\begin{equation*}
    \begin{split}
{E}_d^{img} = f_{\text{encoder}}^{img}({E}_k^{img}), \quad \hat{E}_k^{img} = f_{\text{decoder}}^{img}({E}_d^{img})\\
{E}_d^{text} = f_{\text{encoder}}^{text}({E}_k^{text}), \quad \hat{E}_k^{text} = f_{\text{decoder}}^{text}({E}_d^{text})
\end{split}
\end{equation*}
A $d$-dimensional mask similar to that used in \cite{zhou2024retrievalbaseddisentangledrepresentationlearning} and \cite{formal2021spladesparselexicalexpansion} is created that combines the top $t$ active dimensions of image/text embeddings ($E^{\text{img}}_d$ and $E^{\text{text}}_d$) with the active dimensions from corresponding disentangled sentence embedding ($z_c$) created in training step 2. Thus, the mask captures the dimensions having both modality-specific and shared meaningful features. 
\begin{equation*}
    \begin{small}
E^{\text{img}}_{\text{mask}} = z_c \, \text{OR} \, \text{Top}_t(E^{\text{img}}_d), \quad E^{\text{text}}_{\text{mask}} = z_c \, \text{OR} \, \text{Top}_t(E^{\text{text}}_d)
\end{small}
\end{equation*}
The sparse representations are then obtained by element-wise multiplication:
\begin{equation*}
    \begin{small}
SR_{\text{img}} = E^{\text{img}}_{\text{mask}} \odot E^{\text{img}}_d, \quad SR_{\text{text}} = E^{\text{text}}_{\text{mask}} \odot E^{\text{text}}_d
\end{small}
\end{equation*}
\vspace{-1em}

The loss functions used to optimize the model are:
\begin{asparaitem}
    \item {\textbf{Reconstruction Loss\cite{ng2011sparse}:} Preserves information by reconstructing the original 
$k$-dimensional embeddings for both image and text embeddings:}
\begin{equation*}
    \begin{small}
RL = \left\| {E}^{\text{img}}_k - \hat{E}_k^{text} \right\|_2^2 + \left\| {E}^{\text{text}}_k - \hat{E}_k^{img} \right\|_2^2
\end{small}
\end{equation*}
\vspace{-1em}

\item{\textbf{Contrastive Loss:} We use a contrastive loss similar to that used in \cite{radford2021learningtransferablevisualmodels} between the latent $d$ dimensional image and text embeddings to encourage the similarity between related image-text pairs while pushing apart unrelated pairs:}
\begin{equation*}
    \begin{split}
       CL = - \frac{1}{2N}  
       \left( 
       \sum_{i=1}^{N} \log \frac{\exp(\text{sim}({SR}_{\text{img}}^i, {SR}_{\text{text}}^i))}
       {\sum_{j=1}^{N} \exp(\text{sim}({SR}_{\text{img}}^i, {SR}_{\text{text}}^j))} 
       \right. \\  
       \left. + \sum_{i=1}^{N} \log \frac{\exp(\text{sim}({SR}_{\text{text}}^i, {SR}_{\text{img}}^i))}
       {\sum_{j=1}^{N} \exp(\text{sim}({SR}_{\text{text}}^i, {SR}_{\text{img}}^j))}  
       \right)
   \end{split}
\end{equation*}

\end{asparaitem}
The final loss function combines both: $L = RL + \lambda \cdot CL$\\
The final sparse embedding integrates the multimodal pretrained embeddings'(CLIP/BLIP) rich semantic features with sparse text-based syntactic and semantic cues, creating highly interpretable and informative representations $SR_{clip}$ and $SR_{blip}$.

  \section{Experiments}
\subsection{Experimental Protocol} \label{experiments}
\noindent\textbf{Training Datasets:} We use well-established multi-modal benchmarks, \textit{viz.,} MSCOCO~\cite{lin2015microsoftcococommonobjects} (Train: 118K images, Test: 5K images, each with 4-5 captions across 80 categories) and a subset of Conceptual Captions~\cite{Sharma2018ConceptualCA} (Train: 142K image-text pairs, Test: 20K image-text pairs across 174 labels), to train our models. Our model is trained on these datasets to learn disentangled embeddings of images and texts. These embeddings are then utilized, as described in the \textit{Exclusion-Based Retrieval Task Setting}, to address exclusion queries. To evaluate the effectiveness of our model and compare it against existing retrieval approaches, we construct the \textit{Exclusion Query Evaluation Dataset} using the test sets of MSCOCO and Conceptual Captions.\\

\noindent
\textbf{Exclusion Query Evaluation Dataset:} We construct this dataset using the test set images and labels from two widely used datasets: MSCOCO and Conceptual Captions. Queries are formulated as label pairs $(A,B)$, representing the intent to retrieve images that contain object $A$ while excluding object $B$. To ensure high-quality relevance judgments, we rely strictly on ground-truth object annotations provided by each dataset. For MSCOCO, where images are densely annotated with object instance-level labels across 80 categories, we treat the absence of a label as strong evidence of the corresponding object’s absence in the image. This is supported by MSCOCO’s design objective to provide comprehensive and fine-grained object annotations, which has made it a benchmark dataset for object detection, segmentation, and multi-label recognition tasks over the years~\cite{lin2015microsoftcococommonobjects}. We only select label pairs $(A,B)$ when there are enough images labeled with both $A$ and $B$, and when a sufficient number of images are labeled with $A$ but not with $B$. This cautious filtering helps reduce false positives that may arise from missing labels, especially in partially annotated datasets. By applying this image-level filtering, we ensure that the retained image-query pairs are confidently aligned with the intended exclusion condition. For Conceptual Captions, which provides weaker, web-derived text annotations, we apply the same filtering criteria, but acknowledge that the dataset may exhibit more noise. To account for this, we use Conceptual Captions primarily for scaling and diversity, while MSCOCO serves as our high-quality evaluation benchmark. In total, we identify 3.2K valid exclusion queries from MSCOCO and 20K from Conceptual Captions. For a given query $(A,B)$, the ground-truth set consists of test images labeled with $A$ and not with $B$. This results in 5K aligned image-query pairs for MSCOCO and 20K for Conceptual Captions. Due to multi-object annotations, an image can be associated with multiple exclusion queries, and conversely, a single label pair can correspond to multiple relevant images.

While the dataset is constructed using a straightforward query structure—simple label pairs of the form $(A, B)$—its purpose is to fill a critical gap in the multimodal retrieval literature. To the best of our knowledge, no existing benchmark supports exclusion queries with reliable ground-truth image relevance. Our goal is not only to evaluate the capabilities of current retrieval models in handling even basic exclusion logic but also to establish a foundation for studying this underexplored yet important problem setting. In particular, we aim to demonstrate that disentangled representations offer a compelling and interpretable solution to exclusion-based retrieval, outperforming traditional methods on this specialized task. By providing this dataset, we hope to facilitate systematic evaluation and encourage further research into exclusion-aware retrieval and disentanglement-based modeling.

 \begin{figure*}
 \centering
 \includegraphics[width=1\textwidth]{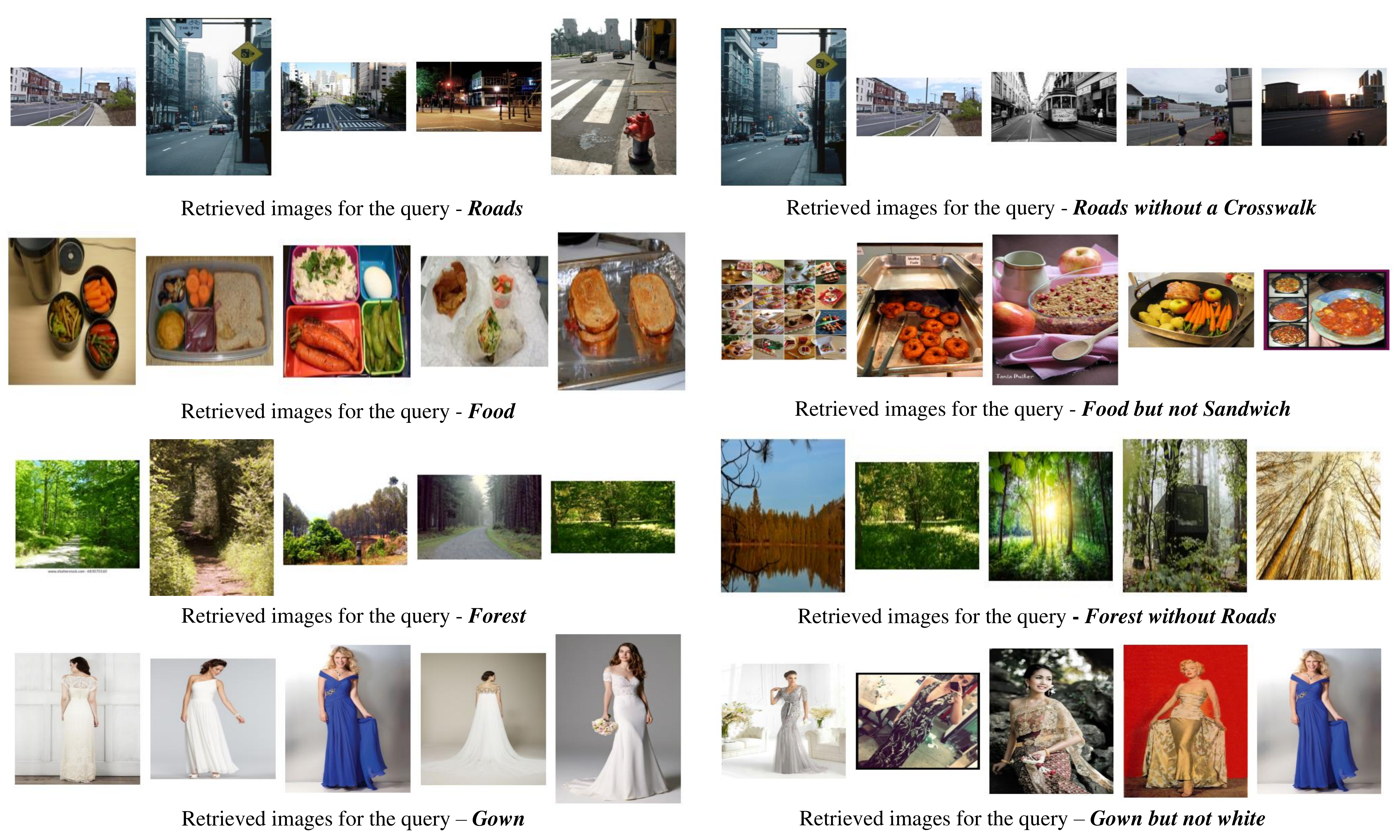} 
 \caption{{Exclusion-based retrieval results using $SR_{clip}$. In each row, the first set of images shows the top retrieved results for a standard query (shown below), while the second set presents the results for the corresponding exclusion-based query, where one object is explicitly excluded from the standard query.}}
 \label{Fig:fig4}
 \vspace{1em}
 \end{figure*}

\vspace{1em}\noindent
\textbf{Exclusion Based Retrieval Task Setting:} Using our disentangled embeddings, we enable controlled retrieval for exclusion queries. For instance, in Figure~\ref{Fig:fig1}, retrieving images for the query \emph{sports but not basketball} follows these steps:
\begin{itemize}
\item \textbf{Dimension Extraction:} We first retrieve the top$-K$ images for a query containing a single label, such as \textit{sports}. From these images, we extract the most active dimensions, denoted as \( D_1 \) by applying a threshold $th$, which selects dimensions contributing to $th\%$ of the embedding’s magnitude. We repeat this process for the label \textit{basketball} to obtain the corresponding dimension set \( D_2 \).    
\item \textbf{Exclusion and Final Retrieval:} To exclude basketball-related features, we subtract set \( D_2 \) from set \( D_1 \), isolating dimensions relevant to sports while eliminating those associated with basketball. Finally, we retrieve the top images based on their highest magnitude in the remaining dimensions.
\end{itemize}
\noindent 
\textbf{Baselines:} We evaluate our proposed representation model against several baseline approaches across different categories:
\\ \noindent
\textbf{1. Multimodal Representations Adapted for Exclusion Retrieval:} Popular vision-language representation models, such as CLIP\cite{radford2021learningtransferablevisualmodels} and BLIP\cite{li2022blipbootstrappinglanguageimagepretraining}, learn joint image-text embeddings via contrastive learning. We adapt these models for exclusion retrieval using two methods:
\begin{itemize}
\item Single-Line Query (\textbf{SLQ}): CLIP/BLIP embeddings are generated for the query \textit{Images of A without B} by treating the query as a single text input.
\item Average Embedding (\textbf{Avg Emb}): For dense models like CLIP and BLIP, we cannot directly apply the Exclusion-Based Retrieval Task Setting used in our approach, since their embeddings are dense and entangled—individual dimensions do not correspond to interpretable concepts. Subtracting or modifying elements in dense vectors often leads to uninterpretable and inconsistent changes. In contrast, our sparse disentangled embeddings explicitly separate concepts across distinct dimensions. Therefore, to adapt CLIP and BLIP for exclusion queries, we follow a method similar to ours by computing the embedding for “A without B” as the average embedding of $A$ minus the average embedding of $B$. This strategy yields the best performance for both models under exclusion evaluation and ensures fairness and consistency in comparison.
\end{itemize}
\noindent
\textbf{2. Retrieval Models:} We use VISTA\cite{zhou-etal-2024-vista}, a state-of-the-art multimodal retrieval model, evaluated with query \textsf{Images of A without B}; and Content-Based Exclusion (CBE)\cite{10.1145/3652583.3657619}, a keyword-based retrieval approach designed for exclusion queries.\\
\noindent
\textbf{3. Disentangled Representation Models:} We evaluate two representative disentangled representation models: VDR~\cite{zhou2024retrievalbaseddisentangledrepresentationlearning}, a sparse representation model that maps visual and textual data into a lexical space, where each dimension corresponds to a specific vocabulary token; and SpLiCE~\cite{Bhalla2024InterpretingCW}, which decomposes dense CLIP embeddings into sparse combinations of 10,000 human-interpretable and semantically meaningful concepts, improving interpretability.

 \begin{table}[t]
   \centering
      \caption{{Average Precision for Image-to-Text and Text-to-Image tasks on MSCOCO and Conceptual Captions datasets. Statistically significant improvements over VDR, Vista, CLIP, $\mathbf{SR_{clip}}$, BLIP and $\mathbf{SR_{blip}}$  are indicated by superscripts 0, 1, 2, 3, 4 and 5 respectively (measured by paired t-Test with 99\% confidence)}}
   \label{table2}
   \resizebox{0.5\textwidth}{!}{
   \begin{tabular}{@{}llcccccc@{}}
     \toprule
     \multicolumn{2}{c}{\textbf{Datasets}} & \multicolumn{3}{c}{\textbf{Image to Text}} & \multicolumn{3}{c}{\textbf{Text to Image}} \\
     \cmidrule(lr){3-8} 
     \multicolumn{2}{c}{} & \multicolumn{1}{c}{\textbf{AP@1}} & \multicolumn{1}{c}{\textbf{AP@5}} & \multicolumn{1}{c}{\textbf{AP@10}}  & \multicolumn{1}{c}{\textbf{AP@1}} & \multicolumn{1}{c}{\textbf{AP@5}} & \multicolumn{1}{c}{\textbf{AP@10}} \\
     \midrule
     & VDR & 0.2896 & 0.1980 & 0.1405 & 0.1607 & 0.0723 & 0.0471 \\
     & Vista & 0.2958 & 0.1969 & 0.1423 & 0.3116 & 0.1131 & 0.0671 \\
     & CLIP & $0.5002^{01}$ & $0.3349^{01}$ & $0.2226^{01}$ & $0.3045^{01}$ & $0.1096^{01}$ & $0.0662^{01}$ \\
     \rowcolor{orange!30}
     \textbf{MSCOCO} & $SR_{clip}$ & $0.4834^{01}$ & $0.3289^{01}$ & $0.2232^{012}$ & $0.3469^{012}$ & $0.1244^{012}$ & $0.0732^{012}$ \\
     & BLIP & $\textbf{0.7864}^{01235}$ & $\textbf{0.5964}^{01235}$ & $\textbf{0.3721 }^{01235}$ &  $\textbf{0.6196}^{01235}$ & $\textbf{0.1707}^{01235}$ & $\textbf{0.0914}^{01235}$ \\
     \rowcolor{cyan!30}
     & $SR_{blip}$ & $0.7490^{0123}$ & $0.5548^{0123}$ & $0.3510^{0123}$ & $0.5836^{0123}$ & $0.1662^{0123}$ & $0.0895^{0123}$  \\
     \midrule
     & VDR & 0.0562 & 0.0254 & 0.0170 & 0.0470 & 0.0223 & 0.0153 \\
     & Vista & 0.0902 & 0.0356 & 0.0227 & 0.1277 & 0.0483 & 0.0300 \\
     \textbf{Conceptual} & CLIP & $0.1569^{01}$ & $0.0600^{01}$ & $0.0369^{01}$  & $0.1444^{01}$ & $0.0568^{01}$ & $0.0356^{01}$ \\
     \rowcolor{orange!30}
     \multicolumn{1}{c}{$\textbf{Captions}$} & $SR_{clip}$ & $0.1212^{01}$ & $0.0505^{01}$ & $0.0324^{01}$ & $0.1409^{01}$ & $0.0586^{012}$ & $0.0375^{012}$  \\
     & BLIP & $\textbf{0.2346}^{01235}$ & $0.0837^{0123}$ & $0.0507^{0123}$  & $0.2356^{0123}$ & $0.0843^{0123}$ & $0.0511^{0123}$  \\
     \rowcolor{cyan!30}
     & $SR_{blip}$ & $0.2243^{0123}$ & $\textbf{0.0842}^{01234}$& $\textbf{0.0516}^{01234}$ & $\textbf{0.2449}^{01234}$ & $\textbf{0.0900}^{01234}$ & $\textbf{0.0546}^{01234}$ \\
     \bottomrule
   \end{tabular}
   }

 \end{table}

\noindent
 \begin{figure*}
 \centering
 \includegraphics[width=1\textwidth]{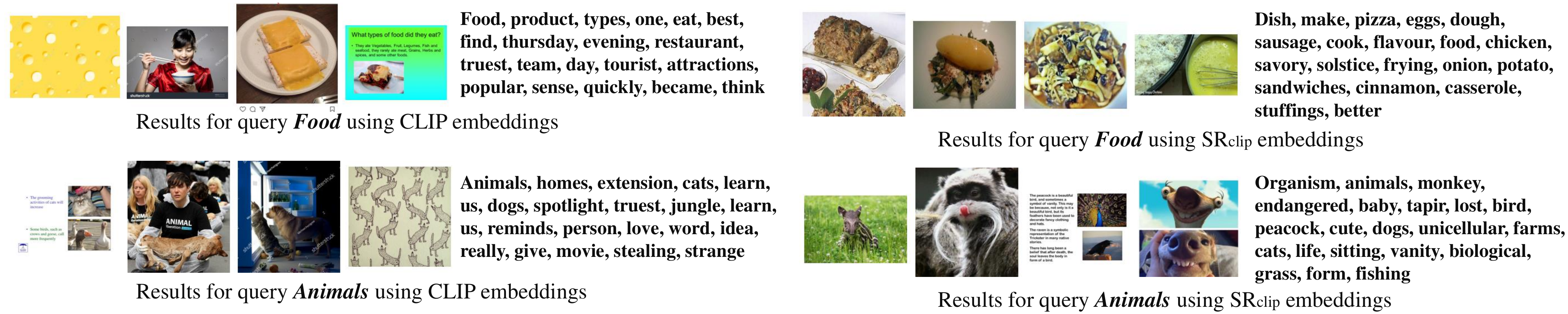} 
 \caption{{Each row illustrates the top retrieved images and the most frequent words from the top retrieved texts on the Conceptual Captions dataset, using CLIP and $SR_{clip}$ embeddings for a given query.}}
 \label{Fig:fig5}
 \vspace{1em}
 \end{figure*}

\subsection{Results and Discussions}
\noindent\textbf{Evaluation on Exclusion Based Queries:} We evaluate retrieval performance using Mean Reciprocal Rank (MRR), Normalized Discounted Cumulative Gain (NDCG), and Average Precision (AP), utilizing label pairs from the \textbf{Exclusion Query Evaluation Dataset} as queries. Our CLIP-based ($SR_{clip}$) and BLIP-based ($SR_{blip}$) embeddings are compared against the baselines outlined in the Baselines Section and results are summarized in Table~\ref{combined_table1}. We note that our method achieves statistically significant outperformance on the MSCOCO dataset over all baselines. However, on the Conceptual Captions dataset, our performance is slightly lower. Further analysis revealed that label inaccuracies in Conceptual Captions -- stemming from the use of automated, nonhuman annotations -- frequently result in correct retrievals being erroneously marked as incorrect by the dataset’s ground truth labels. Examples illustrating these cases are available in our \href{https://anonymous.4open.science/r/retrieval25-C485/supplementary.pdf}{code repository}. Despite this limitation, our models consistently rank among the top two across datasets, demonstrating the robustness and effectiveness of our approach. Figure~\ref{Fig:fig4} showcases a qualitative example that demonstrates the effectiveness of our proposed representations for exclusion-based retrieval. Each row consists of two sets of images: the first set shows the top-ranked results for a standard (non-exclusion) query, while the second set displays the top-ranked images retrieved for the corresponding exclusion-based query. For instance, in the first row, the initial set presents images retrieved for the query Images of Roads, where some results (e.g., the first and third images) include crosswalks. The goal of the exclusion-based query is to retrieve images of roads without crosswalks. The second set illustrates retrieval results using $SR_{clip}$ representations, which effectively filter out images with crosswalks and return only those depicting roads. Similar improvements can be observed in subsequent rows, further highlighting the strength of our approach in handling exclusion queries.\\

\noindent \textbf{Classical Multimodal Retrieval:}
We evaluate our model on standard image-to-text (I2T) and text-to-image (T2I) retrieval tasks using Average Precision scores on the Conceptual Captions and MSCOCO datasets. As presented in Table~\ref{table2}, our BLIP-based model ($SR_{blip}$) excels in T2I retrieval for Conceptual Captions and performs competitively in I2T retrieval, trailing the top-performing model by only a small margin. Additionally, it surpasses the VDR model, which employs a similar sparse architecture, in both retrieval tasks. These results demonstrate that our proposed approach effectively handles exclusion-based retrieval while maintaining strong overall retrieval performance.\\

\noindent
\textbf{Disentanglement:} Recall that our proposed approach effectively disentangles data by activating similar dimensions for semantically related concepts. To illustrate this, Figure~\ref{Fig:fig5} presents examples from the Conceptual Captions dataset, which pairs web images with captions that often lack key visual details. Consequently, retrieval using CLIP embeddings frequently returns mismatched images and captions. In contrast, our $SR_{clip}$ embeddings significantly improve retrieval accuracy by emphasizing contextually relevant features. In Figure~\ref{Fig:fig5}, we compare retrieval results for the queries Food and Animals, displaying both the most similar images and the most frequent words from the top-retrieved captions using CLIP and $SR_{clip}$. Notably, the words retrieved using $SR_{clip}$ align more closely with the intended query, demonstrating the model’s ability to disentangle and organize semantic concepts more effectively. For example, CLIP retrieves unrelated words like find, type, thursday, attractions, evening, etc. for the query Food and retrieves extension, spotlight, truest, learn, us etc. for the query Animal whereas $SR_{clip}$ produces terms that are more relevant to the food and animal domain respectively. 

  \section{Conclusion and Future work}
We propose multimodal representations that are both disentangled and capable of supporting controlled retrieval, with a particular focus on exclusion-based queries. Our results demonstrate that disentangled representations are well-suited for modeling exclusion by effectively isolating and suppressing irrelevant or unwanted concepts. Through this work, we establish the utility of disentanglement as a promising technique for handling exclusion in retrieval. As a direction for future work, the scope and application of disentangled representations can be further explored—extending beyond binary exclusions to support more complex query types, such as multi-label exclusions and conjunctive constraints—thereby advancing the development of more expressive and controllable multimodal retrieval systems.
\begin{acks}
The work of Srikanta Bedathur was partially supported through the DS Chair Professor in Artificial Intelligence. 
\end{acks}
  \bibliographystyle{ACM-Reference-Format}

\begin{thebibliography}{44}


\ifx \showCODEN    \undefined \def \showCODEN     #1{\unskip}     \fi
\ifx \showISBNx    \undefined \def \showISBNx     #1{\unskip}     \fi
\ifx \showISBNxiii \undefined \def \showISBNxiii  #1{\unskip}     \fi
\ifx \showISSN     \undefined \def \showISSN      #1{\unskip}     \fi
\ifx \showLCCN     \undefined \def \showLCCN      #1{\unskip}     \fi
\ifx \shownote     \undefined \def \shownote      #1{#1}          \fi
\ifx \showarticletitle \undefined \def \showarticletitle #1{#1}   \fi
\ifx \showURL      \undefined \def \showURL       {\relax}        \fi
\providecommand\bibfield[2]{#2}
\providecommand\bibinfo[2]{#2}
\providecommand\natexlab[1]{#1}
\providecommand\showeprint[2][]{arXiv:#2}

\bibitem[Bai et~al\mbox{.}(2020)]%
        {bai2020spartermlearningtermbasedsparse}
\bibfield{author}{\bibinfo{person}{Yang Bai}, \bibinfo{person}{Xiaoguang Li},
  \bibinfo{person}{Gang Wang}, \bibinfo{person}{Chaoliang Zhang},
  \bibinfo{person}{Lifeng Shang}, \bibinfo{person}{Jun Xu},
  \bibinfo{person}{Zhaowei Wang}, \bibinfo{person}{Fangshan Wang}, {and}
  \bibinfo{person}{Qun Liu}.} \bibinfo{year}{2020}\natexlab{}.
\newblock \bibinfo{title}{SparTerm: Learning Term-based Sparse Representation
  for Fast Text Retrieval}.
\newblock
\showeprint[arxiv]{2010.00768}~[cs.IR]
\urldef\tempurl%
\url{https://arxiv.org/abs/2010.00768}
\showURL{%
\tempurl}


\bibitem[Bengio et~al\mbox{.}(2013)]%
        {bengio2014representationlearningreviewnew}
\bibfield{author}{\bibinfo{person}{Yoshua Bengio}, \bibinfo{person}{Aaron
  Courville}, {and} \bibinfo{person}{Pascal Vincent}.}
  \bibinfo{year}{2013}\natexlab{}.
\newblock \showarticletitle{Representation Learning: A Review and New
  Perspectives}.
\newblock \bibinfo{journal}{\emph{IEEE Trans. Pattern Anal. Mach. Intell.}}
  \bibinfo{volume}{35}, \bibinfo{number}{8} (\bibinfo{date}{Aug.}
  \bibinfo{year}{2013}), \bibinfo{pages}{1798–1828}.
\newblock
\showISSN{0162-8828}
\href{https://doi.org/10.1109/TPAMI.2013.50}{doi:\nolinkurl{10.1109/TPAMI.2013.50}}


\bibitem[Bhalla et~al\mbox{.}(2024)]%
        {Bhalla2024InterpretingCW}
\bibfield{author}{\bibinfo{person}{Usha Bhalla}, \bibinfo{person}{Alexander~X.
  Oesterling}, \bibinfo{person}{Suraj Srinivas}, \bibinfo{person}{Fl{\'a}vio du
  Pin~Calmon}, {and} \bibinfo{person}{Himabindu Lakkaraju}.}
  \bibinfo{year}{2024}\natexlab{}.
\newblock \showarticletitle{Interpreting CLIP with Sparse Linear Concept
  Embeddings (SpLiCE)}.
\newblock \bibinfo{journal}{\emph{ArXiv}}  \bibinfo{volume}{abs/2402.10376}
  (\bibinfo{year}{2024}).
\newblock
\urldef\tempurl%
\url{https://api.semanticscholar.org/CorpusID:267740469}
\showURL{%
\tempurl}


\bibitem[Bui et~al\mbox{.}(2024)]%
        {bui2024neintellingdontwant}
\bibfield{author}{\bibinfo{person}{Nhat-Tan Bui}, \bibinfo{person}{Dinh-Hieu
  Hoang}, \bibinfo{person}{Quoc-Huy Trinh}, \bibinfo{person}{Minh-Triet Tran},
  \bibinfo{person}{Truong Nguyen}, {and} \bibinfo{person}{Susan Gauch}.}
  \bibinfo{year}{2024}\natexlab{}.
\newblock \bibinfo{title}{NeIn: Telling What You Don't Want}.
\newblock
\showeprint[arxiv]{2409.06481}~[cs.CV]
\urldef\tempurl%
\url{https://arxiv.org/abs/2409.06481}
\showURL{%
\tempurl}


\bibitem[Burgess and Kim(2018)]%
        {3dshapes18}
\bibfield{author}{\bibinfo{person}{Chris Burgess} {and}
  \bibinfo{person}{Hyunjik Kim}.} \bibinfo{year}{2018}\natexlab{}.
\newblock \bibinfo{title}{3D Shapes Dataset}.
\newblock
  \bibinfo{howpublished}{https://github.com/deepmind/3dshapes-dataset/}.
\newblock


\bibitem[Chen et~al\mbox{.}(2018)]%
        {chen2019isolatingsourcesdisentanglementvariational}
\bibfield{author}{\bibinfo{person}{Ricky T.~Q. Chen}, \bibinfo{person}{Xuechen
  Li}, \bibinfo{person}{Roger Grosse}, {and} \bibinfo{person}{David Duvenaud}.}
  \bibinfo{year}{2018}\natexlab{}.
\newblock \showarticletitle{Isolating sources of disentanglement in VAEs}. In
  \bibinfo{booktitle}{\emph{Proceedings of the 32nd International Conference on
  Neural Information Processing Systems}} (Montr\'{e}al, Canada)
  \emph{(\bibinfo{series}{NIPS'18})}. \bibinfo{publisher}{Curran Associates
  Inc.}, \bibinfo{address}{Red Hook, NY, USA}, \bibinfo{pages}{2615–2625}.
\newblock


\bibitem[Formal et~al\mbox{.}(2021a)]%
        {formal2021spladev2sparselexical}
\bibfield{author}{\bibinfo{person}{Thibault Formal}, \bibinfo{person}{Carlos
  Lassance}, \bibinfo{person}{Benjamin Piwowarski}, {and}
  \bibinfo{person}{Stéphane Clinchant}.} \bibinfo{year}{2021}\natexlab{a}.
\newblock \bibinfo{title}{SPLADE v2: Sparse Lexical and Expansion Model for
  Information Retrieval}.
\newblock
\showeprint[arxiv]{2109.10086}~[cs.IR]
\urldef\tempurl%
\url{https://arxiv.org/abs/2109.10086}
\showURL{%
\tempurl}


\bibitem[Formal et~al\mbox{.}(2021b)]%
        {formal2021spladesparselexicalexpansion}
\bibfield{author}{\bibinfo{person}{Thibault Formal}, \bibinfo{person}{Benjamin
  Piwowarski}, {and} \bibinfo{person}{St\'{e}phane Clinchant}.}
  \bibinfo{year}{2021}\natexlab{b}.
\newblock \showarticletitle{SPLADE: Sparse Lexical and Expansion Model for
  First Stage Ranking}. In \bibinfo{booktitle}{\emph{Proceedings of the 44th
  International ACM SIGIR Conference on Research and Development in Information
  Retrieval}} (Virtual Event, Canada) \emph{(\bibinfo{series}{SIGIR '21})}.
  \bibinfo{publisher}{Association for Computing Machinery},
  \bibinfo{address}{New York, NY, USA}, \bibinfo{pages}{2288–2292}.
\newblock
\showISBNx{9781450380379}
\href{https://doi.org/10.1145/3404835.3463098}{doi:\nolinkurl{10.1145/3404835.3463098}}


\bibitem[Garc{\'i}a-Ferrero et~al\mbox{.}(2023)]%
        {garcia-ferrero-etal-2023-dataset}
\bibfield{author}{\bibinfo{person}{Iker Garc{\'i}a-Ferrero},
  \bibinfo{person}{Bego{\~n}a Altuna}, \bibinfo{person}{Javier Alvez},
  \bibinfo{person}{Itziar Gonzalez-Dios}, {and} \bibinfo{person}{German
  Rigau}.} \bibinfo{year}{2023}\natexlab{}.
\newblock \showarticletitle{This is not a Dataset: A Large Negation Benchmark
  to Challenge Large Language Models}. In \bibinfo{booktitle}{\emph{Proceedings
  of the 2023 Conference on Empirical Methods in Natural Language Processing}},
  \bibfield{editor}{\bibinfo{person}{Houda Bouamor}, \bibinfo{person}{Juan
  Pino}, {and} \bibinfo{person}{Kalika Bali}} (Eds.).
  \bibinfo{publisher}{Association for Computational Linguistics},
  \bibinfo{address}{Singapore}, \bibinfo{pages}{8596--8615}.
\newblock
\href{https://doi.org/10.18653/v1/2023.emnlp-main.531}{doi:\nolinkurl{10.18653/v1/2023.emnlp-main.531}}


\bibitem[Greff et~al\mbox{.}(2020)]%
        {Greff2020OnTB}
\bibfield{author}{\bibinfo{person}{Klaus Greff}, \bibinfo{person}{Sjoerd van
  Steenkiste}, {and} \bibinfo{person}{J{\"u}rgen Schmidhuber}.}
  \bibinfo{year}{2020}\natexlab{}.
\newblock \showarticletitle{On the Binding Problem in Artificial Neural
  Networks}.
\newblock \bibinfo{journal}{\emph{ArXiv}}  \bibinfo{volume}{abs/2012.05208}
  (\bibinfo{year}{2020}).
\newblock
\urldef\tempurl%
\url{https://api.semanticscholar.org/CorpusID:228063925}
\showURL{%
\tempurl}


\bibitem[Higgins et~al\mbox{.}(2018)]%
        {higgins2018definitiondisentangledrepresentations}
\bibfield{author}{\bibinfo{person}{Irina Higgins}, \bibinfo{person}{David
  Amos}, \bibinfo{person}{David Pfau}, \bibinfo{person}{Sebastien Racaniere},
  \bibinfo{person}{Loic Matthey}, \bibinfo{person}{Danilo Rezende}, {and}
  \bibinfo{person}{Alexander Lerchner}.} \bibinfo{year}{2018}\natexlab{}.
\newblock \bibinfo{title}{Towards a Definition of Disentangled
  Representations}.
\newblock
\showeprint[arxiv]{1812.02230}~[cs.LG]
\urldef\tempurl%
\url{https://arxiv.org/abs/1812.02230}
\showURL{%
\tempurl}


\bibitem[Higgins et~al\mbox{.}(2017)]%
        {higgins2017betavae}
\bibfield{author}{\bibinfo{person}{Irina Higgins}, \bibinfo{person}{Loic
  Matthey}, \bibinfo{person}{Arka Pal}, \bibinfo{person}{Christopher Burgess},
  \bibinfo{person}{Xavier Glorot}, \bibinfo{person}{Matthew Botvinick},
  \bibinfo{person}{Shakir Mohamed}, {and} \bibinfo{person}{Alexander
  Lerchner}.} \bibinfo{year}{2017}\natexlab{}.
\newblock \showarticletitle{beta-{VAE}: Learning Basic Visual Concepts with a
  Constrained Variational Framework}. In
  \bibinfo{booktitle}{\emph{International Conference on Learning
  Representations}}.
\newblock
\urldef\tempurl%
\url{https://openreview.net/forum?id=Sy2fzU9gl}
\showURL{%
\tempurl}


\bibitem[Huben et~al\mbox{.}(2024)]%
        {huben2024sparse}
\bibfield{author}{\bibinfo{person}{Robert Huben}, \bibinfo{person}{Hoagy
  Cunningham}, \bibinfo{person}{Logan~Riggs Smith}, \bibinfo{person}{Aidan
  Ewart}, {and} \bibinfo{person}{Lee Sharkey}.}
  \bibinfo{year}{2024}\natexlab{}.
\newblock \showarticletitle{Sparse Autoencoders Find Highly Interpretable
  Features in Language Models}. In \bibinfo{booktitle}{\emph{The Twelfth
  International Conference on Learning Representations}}.
\newblock
\urldef\tempurl%
\url{https://openreview.net/forum?id=F76bwRSLeK}
\showURL{%
\tempurl}


\bibitem[Jia et~al\mbox{.}(2023)]%
        {10.1109/TMM.2022.3141267}
\bibfield{author}{\bibinfo{person}{Mengxi Jia}, \bibinfo{person}{Xinhua Cheng},
  \bibinfo{person}{Shijian Lu}, {and} \bibinfo{person}{Jian Zhang}.}
  \bibinfo{year}{2023}\natexlab{}.
\newblock \showarticletitle{Learning Disentangled Representation Implicitly Via
  Transformer for Occluded Person Re-Identification}.
\newblock \bibinfo{journal}{\emph{Trans. Multi.}}  \bibinfo{volume}{25}
  (\bibinfo{date}{Jan.} \bibinfo{year}{2023}), \bibinfo{pages}{1294–1305}.
\newblock
\showISSN{1520-9210}
\href{https://doi.org/10.1109/TMM.2022.3141267}{doi:\nolinkurl{10.1109/TMM.2022.3141267}}


\bibitem[Kim and Mnih(2018)]%
        {kim2019disentanglingfactorising}
\bibfield{author}{\bibinfo{person}{Hyunjik Kim} {and} \bibinfo{person}{Andriy
  Mnih}.} \bibinfo{year}{2018}\natexlab{}.
\newblock \showarticletitle{Disentangling by Factorising}. In
  \bibinfo{booktitle}{\emph{Proceedings of the 35th International Conference on
  Machine Learning}} \emph{(\bibinfo{series}{Proceedings of Machine Learning
  Research}, Vol.~\bibinfo{volume}{80})},
  \bibfield{editor}{\bibinfo{person}{Jennifer Dy} {and}
  \bibinfo{person}{Andreas Krause}} (Eds.). \bibinfo{publisher}{PMLR},
  \bibinfo{pages}{2649--2658}.
\newblock
\urldef\tempurl%
\url{https://proceedings.mlr.press/v80/kim18b.html}
\showURL{%
\tempurl}


\bibitem[Kim et~al\mbox{.}(2021)]%
        {Kim2021LearningDF}
\bibfield{author}{\bibinfo{person}{Minyoung Kim}, \bibinfo{person}{Ricardo
  Guerrero}, {and} \bibinfo{person}{Vladimir Pavlovic}.}
  \bibinfo{year}{2021}\natexlab{}.
\newblock \showarticletitle{Learning Disentangled Factors from Paired Data in
  Cross-Modal Retrieval: An Implicit Identifiable VAE Approach}.
\newblock \bibinfo{journal}{\emph{Proceedings of the 29th ACM International
  Conference on Multimedia}} (\bibinfo{year}{2021}).
\newblock
\urldef\tempurl%
\url{https://api.semanticscholar.org/CorpusID:239011496}
\showURL{%
\tempurl}


\bibitem[Kim et~al\mbox{.}(2019)]%
        {kim2019relevancefactorvaelearning}
\bibfield{author}{\bibinfo{person}{Minyoung Kim}, \bibinfo{person}{Yuting
  Wang}, \bibinfo{person}{Pritish Sahu}, {and} \bibinfo{person}{Vladimir
  Pavlovic}.} \bibinfo{year}{2019}\natexlab{}.
\newblock \bibinfo{title}{Relevance Factor VAE: Learning and Identifying
  Disentangled Factors}.
\newblock
\showeprint[arxiv]{1902.01568}~[cs.LG]
\urldef\tempurl%
\url{https://arxiv.org/abs/1902.01568}
\showURL{%
\tempurl}


\bibitem[Lee and Pavlovic(2021)]%
        {lee2020privateshareddisentangledmultimodalvae}
\bibfield{author}{\bibinfo{person}{Mihee Lee} {and} \bibinfo{person}{Vladimir
  Pavlovic}.} \bibinfo{year}{2021}\natexlab{}.
\newblock \showarticletitle{Private-Shared Disentangled Multimodal VAE for
  Learning of Latent Representations}. \bibinfo{pages}{1692--1700}.
\newblock
\href{https://doi.org/10.1109/CVPRW53098.2021.00185}{doi:\nolinkurl{10.1109/CVPRW53098.2021.00185}}


\bibitem[Li et~al\mbox{.}(2022)]%
        {li2022blipbootstrappinglanguageimagepretraining}
\bibfield{author}{\bibinfo{person}{Junnan Li}, \bibinfo{person}{Dongxu Li},
  \bibinfo{person}{Caiming Xiong}, {and} \bibinfo{person}{Steven Hoi}.}
  \bibinfo{year}{2022}\natexlab{}.
\newblock \bibinfo{title}{BLIP: Bootstrapping Language-Image Pre-training for
  Unified Vision-Language Understanding and Generation}.
\newblock
\showeprint[arxiv]{2201.12086}~[cs.CV]
\urldef\tempurl%
\url{https://arxiv.org/abs/2201.12086}
\showURL{%
\tempurl}


\bibitem[Lin et~al\mbox{.}(2014)]%
        {lin2015microsoftcococommonobjects}
\bibfield{author}{\bibinfo{person}{Tsung-Yi Lin}, \bibinfo{person}{Michael
  Maire}, \bibinfo{person}{Serge Belongie}, \bibinfo{person}{James Hays},
  \bibinfo{person}{Pietro Perona}, \bibinfo{person}{Deva Ramanan},
  \bibinfo{person}{Piotr Doll{\'a}r}, {and} \bibinfo{person}{C.~Lawrence
  Zitnick}.} \bibinfo{year}{2014}\natexlab{}.
\newblock \showarticletitle{Microsoft COCO: Common Objects in Context}. In
  \bibinfo{booktitle}{\emph{Computer Vision -- ECCV 2014}},
  \bibfield{editor}{\bibinfo{person}{David Fleet}, \bibinfo{person}{Tomas
  Pajdla}, \bibinfo{person}{Bernt Schiele}, {and} \bibinfo{person}{Tinne
  Tuytelaars}} (Eds.). \bibinfo{publisher}{Springer International Publishing},
  \bibinfo{address}{Cham}, \bibinfo{pages}{740--755}.
\newblock
\showISBNx{978-3-319-10602-1}


\bibitem[Lu et~al\mbox{.}(2021)]%
        {lu-etal-2021-visualsparta}
\bibfield{author}{\bibinfo{person}{Xiaopeng Lu}, \bibinfo{person}{Tiancheng
  Zhao}, {and} \bibinfo{person}{Kyusong Lee}.} \bibinfo{year}{2021}\natexlab{}.
\newblock \showarticletitle{{V}isual{S}parta: An Embarrassingly Simple Approach
  to Large-scale Text-to-Image Search with Weighted Bag-of-words}. In
  \bibinfo{booktitle}{\emph{Proceedings of the 59th Annual Meeting of the
  Association for Computational Linguistics and the 11th International Joint
  Conference on Natural Language Processing (Volume 1: Long Papers)}},
  \bibfield{editor}{\bibinfo{person}{Chengqing Zong}, \bibinfo{person}{Fei
  Xia}, \bibinfo{person}{Wenjie Li}, {and} \bibinfo{person}{Roberto Navigli}}
  (Eds.). \bibinfo{publisher}{Association for Computational Linguistics},
  \bibinfo{address}{Online}, \bibinfo{pages}{5020--5029}.
\newblock
\href{https://doi.org/10.18653/v1/2021.acl-long.389}{doi:\nolinkurl{10.18653/v1/2021.acl-long.389}}


\bibitem[Luo et~al\mbox{.}(2023)]%
        {Luo_2023_ICCV}
\bibfield{author}{\bibinfo{person}{Ziyang Luo}, \bibinfo{person}{Pu Zhao},
  \bibinfo{person}{Can Xu}, \bibinfo{person}{Xiubo Geng}, \bibinfo{person}{Tao
  Shen}, \bibinfo{person}{Chongyang Tao}, \bibinfo{person}{Jing Ma},
  \bibinfo{person}{Qingwei Lin}, {and} \bibinfo{person}{Daxin Jiang}.}
  \bibinfo{year}{2023}\natexlab{}.
\newblock \showarticletitle{LexLIP: Lexicon-Bottlenecked Language-Image
  Pre-Training for Large-Scale Image-Text Sparse Retrieval}. In
  \bibinfo{booktitle}{\emph{Proceedings of the IEEE/CVF International
  Conference on Computer Vision (ICCV)}}. \bibinfo{pages}{11206--11217}.
\newblock


\bibitem[Ma et~al\mbox{.}(2019)]%
        {ma2019learningdisentangledrepresentationsrecommendation}
\bibfield{author}{\bibinfo{person}{Jianxin Ma}, \bibinfo{person}{Chang Zhou},
  \bibinfo{person}{Peng Cui}, \bibinfo{person}{Hongxia Yang}, {and}
  \bibinfo{person}{Wenwu Zhu}.} \bibinfo{year}{2019}\natexlab{}.
\newblock \bibinfo{title}{Learning Disentangled Representations for
  Recommendation}.
\newblock
\showeprint[arxiv]{1910.14238}~[cs.LG]
\urldef\tempurl%
\url{https://arxiv.org/abs/1910.14238}
\showURL{%
\tempurl}


\bibitem[Malaviya et~al\mbox{.}(2023)]%
        {malaviya-etal-2023-quest}
\bibfield{author}{\bibinfo{person}{Chaitanya Malaviya}, \bibinfo{person}{Peter
  Shaw}, \bibinfo{person}{Ming-Wei Chang}, \bibinfo{person}{Kenton Lee}, {and}
  \bibinfo{person}{Kristina Toutanova}.} \bibinfo{year}{2023}\natexlab{}.
\newblock \showarticletitle{{QUEST}: A Retrieval Dataset of Entity-Seeking
  Queries with Implicit Set Operations}. In
  \bibinfo{booktitle}{\emph{Proceedings of the 61st Annual Meeting of the
  Association for Computational Linguistics (Volume 1: Long Papers)}},
  \bibfield{editor}{\bibinfo{person}{Anna Rogers}, \bibinfo{person}{Jordan
  Boyd-Graber}, {and} \bibinfo{person}{Naoaki Okazaki}} (Eds.).
  \bibinfo{publisher}{Association for Computational Linguistics},
  \bibinfo{address}{Toronto, Canada}, \bibinfo{pages}{14032--14047}.
\newblock
\href{https://doi.org/10.18653/v1/2023.acl-long.784}{doi:\nolinkurl{10.18653/v1/2023.acl-long.784}}


\bibitem[Mondal et~al\mbox{.}(2022)]%
        {article}
\bibfield{author}{\bibinfo{person}{Arnab Mondal}, \bibinfo{person}{Ajay
  Sailopal}, \bibinfo{person}{Parag Singla}, {and} \bibinfo{person}{A~P
  Prathosh}.} \bibinfo{year}{2022}\natexlab{}.
\newblock \showarticletitle{SSDMM-VAE: variational multi-modal disentangled
  representation learning}.
\newblock \bibinfo{journal}{\emph{Applied Intelligence}}  \bibinfo{volume}{53}
  (\bibinfo{date}{07} \bibinfo{year}{2022}).
\newblock
\href{https://doi.org/10.1007/s10489-022-03936-z}{doi:\nolinkurl{10.1007/s10489-022-03936-z}}


\bibitem[Ng et~al\mbox{.}({[n.\,d.]})]%
        {ng2011sparse}
\bibfield{author}{\bibinfo{person}{Andrew Ng} {et~al\mbox{.}}}
  \bibinfo{year}{[n.\,d.]}\natexlab{}.
\newblock \showarticletitle{Sparse autoencoder}.
\newblock  (\bibinfo{year}{[n.\,d.]}).
\newblock


\bibitem[Nguyen et~al\mbox{.}(2024)]%
        {nguyen2024multimodallearnedsparseretrieval}
\bibfield{author}{\bibinfo{person}{Thong Nguyen}, \bibinfo{person}{Mariya
  Hendriksen}, \bibinfo{person}{Andrew Yates}, {and} \bibinfo{person}{Maarten
  de Rijke}.} \bibinfo{year}{2024}\natexlab{}.
\newblock \bibinfo{title}{Multimodal Learned Sparse Retrieval with
  Probabilistic Expansion Control}.
\newblock
\showeprint[arxiv]{2402.17535}~[cs.IR]
\urldef\tempurl%
\url{https://arxiv.org/abs/2402.17535}
\showURL{%
\tempurl}


\bibitem[O'Neill et~al\mbox{.}(2024)]%
        {o'neill2024towards}
\bibfield{author}{\bibinfo{person}{Charles O'Neill}, \bibinfo{person}{Christine
  Ye}, \bibinfo{person}{Kartheik~G. Iyer}, {and} \bibinfo{person}{John~F Wu}.}
  \bibinfo{year}{2024}\natexlab{}.
\newblock \showarticletitle{Towards Interpretable Scientific Foundation Models:
  Sparse Autoencoders for Disentangling Dense Embeddings of Scientific
  Concepts}. In \bibinfo{booktitle}{\emph{Neurips 2024 Workshop Foundation
  Models for Science: Progress, Opportunities, and Challenges}}.
\newblock
\urldef\tempurl%
\url{https://openreview.net/forum?id=mPq3R6jdtD}
\showURL{%
\tempurl}


\bibitem[Pennington et~al\mbox{.}(2014)]%
        {pennington-etal-2014-glove}
\bibfield{author}{\bibinfo{person}{Jeffrey Pennington},
  \bibinfo{person}{Richard Socher}, {and} \bibinfo{person}{Christopher
  Manning}.} \bibinfo{year}{2014}\natexlab{}.
\newblock \showarticletitle{{G}lo{V}e: Global Vectors for Word Representation}.
  In \bibinfo{booktitle}{\emph{Proceedings of the 2014 Conference on Empirical
  Methods in Natural Language Processing ({EMNLP})}},
  \bibfield{editor}{\bibinfo{person}{Alessandro Moschitti},
  \bibinfo{person}{Bo~Pang}, {and} \bibinfo{person}{Walter Daelemans}} (Eds.).
  \bibinfo{publisher}{Association for Computational Linguistics},
  \bibinfo{address}{Doha, Qatar}, \bibinfo{pages}{1532--1543}.
\newblock
\href{https://doi.org/10.3115/v1/D14-1162}{doi:\nolinkurl{10.3115/v1/D14-1162}}


\bibitem[Pizzati et~al\mbox{.}(2020)]%
        {pizzati2020modelbasedocclusiondisentanglementimagetoimage}
\bibfield{author}{\bibinfo{person}{Fabio Pizzati}, \bibinfo{person}{Pietro
  Cerri}, {and} \bibinfo{person}{Raoul de Charette}.}
  \bibinfo{year}{2020}\natexlab{}.
\newblock \bibinfo{title}{Model-based occlusion disentanglement for
  image-to-image translation}.
\newblock
\showeprint[arxiv]{2004.01071}~[cs.CV]
\urldef\tempurl%
\url{https://arxiv.org/abs/2004.01071}
\showURL{%
\tempurl}


\bibitem[Radford et~al\mbox{.}(2021)]%
        {radford2021learningtransferablevisualmodels}
\bibfield{author}{\bibinfo{person}{Alec Radford}, \bibinfo{person}{Jong~Wook
  Kim}, \bibinfo{person}{Chris Hallacy}, \bibinfo{person}{Aditya Ramesh},
  \bibinfo{person}{Gabriel Goh}, \bibinfo{person}{Sandhini Agarwal},
  \bibinfo{person}{Girish Sastry}, \bibinfo{person}{Amanda Askell},
  \bibinfo{person}{Pamela Mishkin}, \bibinfo{person}{Jack Clark},
  \bibinfo{person}{Gretchen Krueger}, {and} \bibinfo{person}{Ilya Sutskever}.}
  \bibinfo{year}{2021}\natexlab{}.
\newblock \bibinfo{title}{Learning Transferable Visual Models From Natural
  Language Supervision}.
\newblock
\showeprint[arxiv]{2103.00020}~[cs.CV]
\urldef\tempurl%
\url{https://arxiv.org/abs/2103.00020}
\showURL{%
\tempurl}


\bibitem[Ravichander et~al\mbox{.}(2022)]%
        {Ravichander2022CONDAQAAC}
\bibfield{author}{\bibinfo{person}{Abhilasha Ravichander},
  \bibinfo{person}{Matt Gardner}, {and} \bibinfo{person}{Ana Marasovi{\'c}}.}
  \bibinfo{year}{2022}\natexlab{}.
\newblock \showarticletitle{CONDAQA: A Contrastive Reading Comprehension
  Dataset for Reasoning about Negation}.
\newblock \bibinfo{journal}{\emph{ArXiv}}  \bibinfo{volume}{abs/2211.00295}
  (\bibinfo{year}{2022}).
\newblock
\urldef\tempurl%
\url{https://api.semanticscholar.org/CorpusID:253244137}
\showURL{%
\tempurl}


\bibitem[Sharma et~al\mbox{.}(2018)]%
        {Sharma2018ConceptualCA}
\bibfield{author}{\bibinfo{person}{Piyush Sharma}, \bibinfo{person}{Nan Ding},
  \bibinfo{person}{Sebastian Goodman}, {and} \bibinfo{person}{Radu Soricut}.}
  \bibinfo{year}{2018}\natexlab{}.
\newblock \showarticletitle{Conceptual Captions: A Cleaned, Hypernymed, Image
  Alt-text Dataset For Automatic Image Captioning}. In
  \bibinfo{booktitle}{\emph{Annual Meeting of the Association for Computational
  Linguistics}}.
\newblock
\urldef\tempurl%
\url{https://api.semanticscholar.org/CorpusID:51876975}
\showURL{%
\tempurl}


\bibitem[Singh et~al\mbox{.}(2024)]%
        {singh2024learnnosay}
\bibfield{author}{\bibinfo{person}{Jaisidh Singh}, \bibinfo{person}{Ishaan
  Shrivastava}, \bibinfo{person}{Mayank Vatsa}, \bibinfo{person}{Richa Singh},
  {and} \bibinfo{person}{Aparna Bharati}.} \bibinfo{year}{2024}\natexlab{}.
\newblock \bibinfo{title}{Learn "No" to Say "Yes" Better: Improving
  Vision-Language Models via Negations}.
\newblock
\showeprint[arxiv]{2403.20312}~[cs.CV]
\urldef\tempurl%
\url{https://arxiv.org/abs/2403.20312}
\showURL{%
\tempurl}


\bibitem[Su et~al\mbox{.}(2025)]%
        {Su_2025}
\bibfield{author}{\bibinfo{person}{Yixin Su}, \bibinfo{person}{Wei Jiang},
  \bibinfo{person}{Fangquan Lin}, \bibinfo{person}{Cheng Yang},
  \bibinfo{person}{Sarah~M. Erfani}, \bibinfo{person}{Junhao Gan},
  \bibinfo{person}{Yunxiang Zhao}, \bibinfo{person}{Ruixuan Li}, {and}
  \bibinfo{person}{Rui Zhang}.} \bibinfo{year}{2025}\natexlab{}.
\newblock \showarticletitle{Intrinsic and Extrinsic Factor Disentanglement for
  Recommendation in Various Context Scenarios}.
\newblock \bibinfo{journal}{\emph{ACM Transactions on Information Systems}}
  (\bibinfo{date}{March} \bibinfo{year}{2025}).
\newblock
\showISSN{1558-2868}
\href{https://doi.org/10.1145/3722553}{doi:\nolinkurl{10.1145/3722553}}


\bibitem[Subramanian et~al\mbox{.}(2017)]%
        {Subramanian2017SPINESI}
\bibfield{author}{\bibinfo{person}{Anant Subramanian}, \bibinfo{person}{Danish
  Pruthi}, \bibinfo{person}{Harsh Jhamtani}, \bibinfo{person}{Taylor
  Berg-Kirkpatrick}, {and} \bibinfo{person}{Eduard~H. Hovy}.}
  \bibinfo{year}{2017}\natexlab{}.
\newblock \showarticletitle{SPINE: SParse Interpretable Neural Embeddings}.
\newblock \bibinfo{journal}{\emph{ArXiv}}  \bibinfo{volume}{abs/1711.08792}
  (\bibinfo{year}{2017}).
\newblock
\urldef\tempurl%
\url{https://api.semanticscholar.org/CorpusID:19143983}
\showURL{%
\tempurl}


\bibitem[Wang et~al\mbox{.}(2024a)]%
        {wang2024disentangledrepresentationlearning}
\bibfield{author}{\bibinfo{person}{Xin Wang}, \bibinfo{person}{Hong Chen},
  \bibinfo{person}{Si'ao Tang}, \bibinfo{person}{Zihao Wu}, {and}
  \bibinfo{person}{Wenwu Zhu}.} \bibinfo{year}{2024}\natexlab{a}.
\newblock \bibinfo{title}{Disentangled Representation Learning}.
\newblock
\showeprint[arxiv]{2211.11695}~[cs.LG]
\urldef\tempurl%
\url{https://arxiv.org/abs/2211.11695}
\showURL{%
\tempurl}


\bibitem[Wang et~al\mbox{.}(2024b)]%
        {wang2024disentangled}
\bibfield{author}{\bibinfo{person}{Xin Wang}, \bibinfo{person}{Hong Chen},
  \bibinfo{person}{Zihao Wu}, \bibinfo{person}{Wenwu Zhu}, {et~al\mbox{.}}}
  \bibinfo{year}{2024}\natexlab{b}.
\newblock \showarticletitle{Disentangled representation learning}.
\newblock \bibinfo{journal}{\emph{IEEE Transactions on Pattern Analysis and
  Machine Intelligence}} (\bibinfo{year}{2024}).
\newblock


\bibitem[Weller et~al\mbox{.}(2024)]%
        {weller-etal-2024-nevir}
\bibfield{author}{\bibinfo{person}{Orion Weller}, \bibinfo{person}{Dawn
  Lawrie}, {and} \bibinfo{person}{Benjamin Van~Durme}.}
  \bibinfo{year}{2024}\natexlab{}.
\newblock \showarticletitle{{N}ev{IR}: Negation in Neural Information
  Retrieval}. In \bibinfo{booktitle}{\emph{Proceedings of the 18th Conference
  of the European Chapter of the Association for Computational Linguistics
  (Volume 1: Long Papers)}}, \bibfield{editor}{\bibinfo{person}{Yvette Graham}
  {and} \bibinfo{person}{Matthew Purver}} (Eds.).
  \bibinfo{publisher}{Association for Computational Linguistics},
  \bibinfo{address}{St. Julian{'}s, Malta}, \bibinfo{pages}{2274--2287}.
\newblock
\urldef\tempurl%
\url{https://aclanthology.org/2024.eacl-long.139/}
\showURL{%
\tempurl}


\bibitem[Yoshikawa and Tajima(2024)]%
        {10.1145/3652583.3657619}
\bibfield{author}{\bibinfo{person}{Eisaku Yoshikawa} {and}
  \bibinfo{person}{Keishi Tajima}.} \bibinfo{year}{2024}\natexlab{}.
\newblock \showarticletitle{Content-Based Exclusion Queries in Keyword-Based
  Image Retrieval}. In \bibinfo{booktitle}{\emph{Proceedings of the 2024
  International Conference on Multimedia Retrieval}} (Phuket, Thailand)
  \emph{(\bibinfo{series}{ICMR '24})}. \bibinfo{publisher}{Association for
  Computing Machinery}, \bibinfo{address}{New York, NY, USA},
  \bibinfo{pages}{1145–1149}.
\newblock
\showISBNx{9798400706196}
\href{https://doi.org/10.1145/3652583.3657619}{doi:\nolinkurl{10.1145/3652583.3657619}}


\bibitem[Zhang et~al\mbox{.}(2024a)]%
        {zhang2024chooseneeddisentangledrepresentation}
\bibfield{author}{\bibinfo{person}{Boqiang Zhang}, \bibinfo{person}{Hongtao
  Xie}, \bibinfo{person}{Zuan Gao}, {and} \bibinfo{person}{Yuxin Wang}.}
  \bibinfo{year}{2024}\natexlab{a}.
\newblock \bibinfo{title}{Choose What You Need: Disentangled Representation
  Learning for Scene Text Recognition, Removal and Editing}.
\newblock
\showeprint[arxiv]{2405.04377}~[cs.CV]
\urldef\tempurl%
\url{https://arxiv.org/abs/2405.04377}
\showURL{%
\tempurl}


\bibitem[Zhang et~al\mbox{.}(2024b)]%
        {zhang2024excluirexclusionaryneuralinformation}
\bibfield{author}{\bibinfo{person}{Wenhao Zhang}, \bibinfo{person}{Mengqi
  Zhang}, \bibinfo{person}{Shiguang Wu}, \bibinfo{person}{Jiahuan Pei},
  \bibinfo{person}{Zhaochun Ren}, \bibinfo{person}{Maarten de Rijke},
  \bibinfo{person}{Zhumin Chen}, {and} \bibinfo{person}{Pengjie Ren}.}
  \bibinfo{year}{2024}\natexlab{b}.
\newblock \bibinfo{title}{ExcluIR: Exclusionary Neural Information Retrieval}.
\newblock
\showeprint[arxiv]{2404.17288}~[cs.IR]
\urldef\tempurl%
\url{https://arxiv.org/abs/2404.17288}
\showURL{%
\tempurl}


\bibitem[Zhou et~al\mbox{.}(2024a)]%
        {zhou2024retrievalbaseddisentangledrepresentationlearning}
\bibfield{author}{\bibinfo{person}{Jiawei Zhou}, \bibinfo{person}{Xiaoguang
  Li}, \bibinfo{person}{Lifeng Shang}, \bibinfo{person}{Xin Jiang},
  \bibinfo{person}{Qun Liu}, {and} \bibinfo{person}{Lei Chen}.}
  \bibinfo{year}{2024}\natexlab{a}.
\newblock \bibinfo{title}{Retrieval-based Disentangled Representation Learning
  with Natural Language Supervision}.
\newblock
\showeprint[arxiv]{2212.07699}~[cs.CL]
\urldef\tempurl%
\url{https://arxiv.org/abs/2212.07699}
\showURL{%
\tempurl}


\bibitem[Zhou et~al\mbox{.}(2024b)]%
        {zhou-etal-2024-vista}
\bibfield{author}{\bibinfo{person}{Junjie Zhou}, \bibinfo{person}{Zheng Liu},
  \bibinfo{person}{Shitao Xiao}, \bibinfo{person}{Bo Zhao}, {and}
  \bibinfo{person}{Yongping Xiong}.} \bibinfo{year}{2024}\natexlab{b}.
\newblock \showarticletitle{{VISTA}: Visualized Text Embedding For Universal
  Multi-Modal Retrieval}. In \bibinfo{booktitle}{\emph{Proceedings of the 62nd
  Annual Meeting of the Association for Computational Linguistics (Volume 1:
  Long Papers)}}, \bibfield{editor}{\bibinfo{person}{Lun-Wei Ku},
  \bibinfo{person}{Andre Martins}, {and} \bibinfo{person}{Vivek Srikumar}}
  (Eds.). \bibinfo{publisher}{Association for Computational Linguistics},
  \bibinfo{address}{Bangkok, Thailand}, \bibinfo{pages}{3185--3200}.
\newblock
\href{https://doi.org/10.18653/v1/2024.acl-long.175}{doi:\nolinkurl{10.18653/v1/2024.acl-long.175}}


\end{thebibliography}

\end{document}